# Adaptive Paired-Comparison Method for Subjective Video Quality Assessment on Mobile Devices

Katherine Storrs, Sebastiaan Van Leuven, Steve Kojder, Lucas Theis, Ferenc Huszár
Twitter
20 Air Street, W1B 5AG, London
United Kingdom

*Abstract*—To effectively evaluate subjective visual quality in weakly-controlled environments, we propose an Adaptive Paired Comparison method based on particle filtering. As our approach requires each sample to be rated only once, the test time compared to regular paired comparison can be reduced. The method works with non-experts and improves reliability compared to MOS and DS-MOS methods.

## I. Introduction

Understanding the visual quality of video generated by a production encoding backend is important to improve the visual aspect of users' perceived Quality of Experience (QoE). Unfortunately, objective measures of image fidelity such as PSNR or SSIM [1] often do not accurately predict subjective visual quality. Additionally, standardized subjective quality measurement methodologies tend to provide a single quality metric on a per-video basis. Quality measured on a per-video basis can be undesirable when evaluating large datasets, which try to represent all types of content, since different content ingested in the same pipeline results in different perceived visual quality. This is the case for static and dynamic encoding settings, such as Netflix' per-title encoding [2]. Therefore, the quality value of an encoding system based on per-video quality values tends to have a large confidence interval and may mis-represent the actual perceived visual quality.

Understanding users' QoE also requires taking their viewing environment into account. While existing subjective quality methods [3] assume a controlled environment, online service customers shift away from these classic viewing environments, especially in the case of mobile users. Therefore, subjective visual quality under weakly-controlled conditions is becoming more important in our industry. We have identified that, for a large set of video samples, both Single Stimulus Mean Opinion Score (MOS) and Double stimulus MOS (DS-MOS) did not meet our requirements in terms of accuracy and consistency between experiments. Expert viewing, such as subjective assessment of video quality using Expert Viewing Protocol [4] could provide a solution, but requires a number of experts, which might not be readily available in companies. To allow for a large number of experiments, it is important to have a reliable methodology that can be performed easily by non-experts.

To overcome these problems, we designed a methodology based on paired comparison testing, where test samples are pairwise rated against reference quality samples. To reduce the test time of an exhaustive paired comparison, each sample under test is rated only once, to arrive at an estimate of the subjective quality of an encoding variant, across individual samples. In order to reduce the number of experimental trials, a particle filtering procedure is used to adaptively select the quality level of the reference sample based on the rater's previous responses. This methodology efficiently evaluates a large dataset with high accuracy and reproducible results. While we believe the applications for this Adaptive Paired-Comparison (APC) method are broader than mobile video, we focus on mobile device, as this is our core business.

In this paper, we discuss paired comparison tasks and describe our APC method. In the results section, we will compare our APC methodology with MOS and DS-MOS evaluations performed for the same videos, in weakly-controlled environments on mobile devices.

## II. Paired-Comparison Tasks

In experimental psychology there is a long history of measuring perception using "*n*-alternative forced choice" methods such as paired-comparison tasks to minimise bias and improve reliability [5]. In a two-interval paired-comparison video quality task, two versions of a video clip are played one after the other, and the rater reports whether the first or the second clip had higher quality. One of the videos is encoded at the settings one wishes to measure the quality of (the "standard" video), and the other is encoded at a known quality chosen from a predefined quality scale (the "reference" video). This *relative* judgement has the advantages over single-stimulus rating tasks such as MOS or DS-MOS of (a) not requiring observers to agree on the meanings of numerical or verbal quality ratings, and (b) being less susceptible to overall changes in quality due to e.g. changes in viewing conditions.

## A. Adaptive Stimulus Choice

A major obstacle to using paired-comparison tasks is that it is usually prohibitively time-consuming to exhaustively present all standard videos paired with all possible reference videos. In scientific research, various methods have been devised to get high-confidence quality estimates using just a small subset of possible pairwise comparisons, by "adaptively" choosing which pair to present on the next trial, based on the rater's response(s) on the previous trial(s).

*"Staircase" Methods:* The simplest adaptive methods are "up-down" or "staircase" methods [5]. These involve progressively increasing or decreasing the quality of the reference video shown across trials, based on the rater's response on the previous trial(s). For example, the reference video quality may begin at the highest possible level, and reduce by one level each trial for so long as the rater reports that the reference video looked better than the standard video. Once the rater encounters a trial on which they prefer the standard video, the quality of the reference video is increased by one level. The procedure continues in this manner for as many trials as one has example standard videos, with the quality level of the reference video forming a zig-zagging "staircase" which eventually converges on the quality level which is neither better nor worse than the standard video. This is the "point of subjective equality" (PSE), which we take as an estimate of the quality level of that standard variant being tested. Staircase methods have the virtue of being simple, but are not maximally efficient [6]. This can be seen in the simulation in Figure 1, where the mean squared error in estimating the perceived quality for one observer is shown as a function of numbers of trials.

*Parameter Estimation Methods:* A more sophisticated class of adaptive methods directly selects the next stimulus in order to minimise expected uncertainty about the parameters of an underlying psychometric function. The psychometric function, for a particular standard test variant, relates the values of the reference quality scale to the probability of a rater preferring that reference level to the standard variant. It can be estimated by fitting a sigmoidal function to the rater's response data (reference level shown vs. proportion of "reference level better" responses), and its midpoint taken as the PSE [5]. Parameter estimation adaptive methods can yield higher-quality data (i.e. lower uncertainty in the PSE estimate) and/or faster experiments consisting of fewer trials [6]. Below we outline an efficient Bayesian active learning procedure based on particle filtering, which selects reference levels to present in order to minimise uncertainty about the midpoint of the psychometric function.

## III. BAYESIAN ACTIVE LEARNING PROCEDURE FOR PARAMETER ESTIMATION

Our goal is to infer from a dataset of binary preference judgments the quality $q$ of the standard variant being tested, where $q$ is a value on our 50-point reference quality scale (see following section for a description of the reference scale). We assume that a logistic psychometric function relates reference quality to preferred stimulus, with midpoint $q$ to be estimated, and a fixed slope. We use a weighted particle filtering simulation to approximate the posterior distribution of $q$, as in [7].

Before observing any data we initially sample 225 particles from a uniform prior, and initialise their weights uniformly. The particles and their associated weights represent our current approximate posterior. After observing a new datapoint we update the weights by multiplying them with the likelihood - the probability of the new datapoint given the particle - then normalising so the weights sum to one. We select the next reference stimulus given the dataset so far, based on a Bayesian Active Learning by Disagreement (BALD criterion). We measure the mutual information between the subject's response in the next trial, and the parameters. This acquisition function is computed for all 50 possible reference levels and the reference level with the maximum acquisition value is chosen as the reference stimulus on the next trial.

In simulations, our active learning procedure reaches more closely estimates the true parameter value of a simulated observer when using small numbers of trials (fewer than 100) than does either random reference level selection or a heuristic "staircase" procedure (see Figure 1).

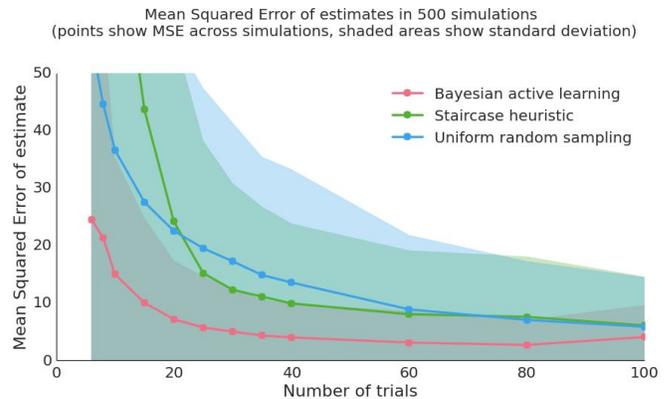

Figure 1. Mean squared error of the estimated perceived quality for one simulated observer.

## A. Creation of a Perceptually Linear Reference Scale

The APC method requires that for every test video, the same video clip is available encoded at all locations on a scale which increases in perceived quality monotonically, and ideally linearly. We constructed this reference scale by first creating an approximate scale based on objective quality (PSNR values) and then adjusting it via human experiments to get closer to perceptual linearity.

Objective PSNR values were first calculated for a set of 100 sample videos encoded at seven different resolutions, and all integer CRF values between 12-52.

1. We then sampled the average bitrate at 50 logarithmically spaced points, taking at each average bitrate the combination of CRF and resolution which had yielded the highest PSNR. This can be thought of as sampling along the convex hull of the seven per-resolution PSNR curves (Figure 2, top panel).
2. A human experiment was run in which 22 raters judged the relative quality of pairs of videos sampled from combinations of points in this initial scale. Gradient descent was used to infer the relative perceptual quality of each reference level. Non-linearities in the estimated perceptual quality function indicate parts of the reference scale where perceived quality changes between levels in a non-uniform fashion. This estimated function was resampled so that steps in the adjusted reference scale are approximately perceptually linear (Figure 2, bottom panel).

A final validation experiment was run in which pairs from the adjusted reference scale were shown to 22 human raters, and the perceptual quality of each level was again estimated. The linearity of the scale improved (normalised MSE of the initial scale relative to perfect linearity = 1.70, and of the adjusted scale = 0.92), and the adjusted scale is used in all subsequent APC testing.

IV. COMPARISON OF APC TO MOS AND DS-MOS IN HUMAN EXPERIMENTS

A. *Stimulus Video Set*

Thirty ten-second clips from high quality original videos, with diverse and difficult-to-compress content (complex textures, camera motion and/or object motion), were selected from sources such as the open source short film *Tears of Steel* (www.mango.blender.org), the CDVL database (www.cdvl.org) and the Netflix open source test footage *El Fuente*.

B. *Experimental Design*

To assess the precision and reliability of APC compared to MOS and DS-MOS, we used each of the methods to measure the subjective visual quality of the same videos encoded at low (320x180@256kbps), medium (640x260@768kbps) and high quality (1280x720@2.5mpsb) rate points. Typical production encoding settings for mobile video were used: H.264/AVC main profile, 3 sec segments/GOPs, 5 B-frames, and fixed CRF (28) with vbv-maxrate and vbv-bufsize equal to the bit rate.

Raters viewed videos in full-screen presentation on iPhone 6s devices via a custom-written application. For each method, two testing sessions were conducted, on separate days, involving the following standard variants:

1. Session 1: 180p and 360p resolution
2. Session 2: 360p and 720p resolution

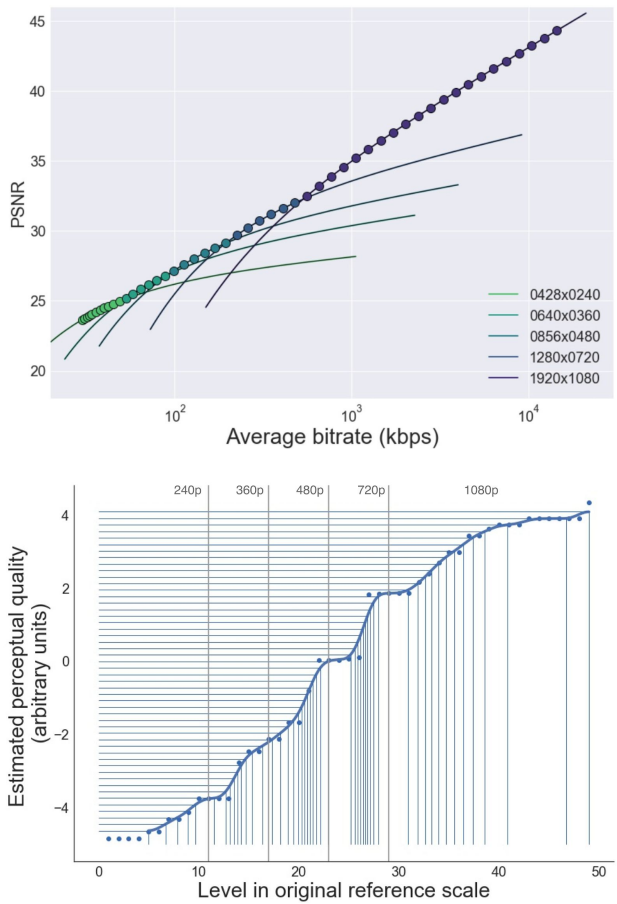

Figure 2. (Top) Construction of an initial reference scale, using the convex hull of the RD curves of five different resolutions. (Bottom) Resampling the initial reference scale (*x*-axis) to linearly increase in perceptual quality (*y*-axis), based on the estimated perceptual quality (blue curve).

For each method, we ran two tests. In a first test we evaluate the quality of the 180p and 360p variants. In a second tests, the same 360p samples and 720p samples were evaluated. This allows us to use the 360p samples to assess test-retest reliability, since a known problem in MOS testing is that quality ratings are biased by the quality of other videos shown during the same or recent sessions.

*Mean Opinion Score method:* On each trial, one ten-second clip from one of the two standard variants was presented, and the rater judged its quality on a 5-point scale labelled "Bad," "Poor," "Fair," "Good," and "Excellent." This led to a total of 60 trials per session, which were presented in a different random order for each rater. Twenty-two raters took part in the first session, and 22 in the second session (20 raters took part in both sessions). Responses from each rater were screened to check for random or non-discriminating responses, and all data were included in the analysis.

*Double-Stimulus Mean Opinion Score method:* On each trial, a pair of ten-second clips was presented. The first video was always a high quality, minimally impaired version of the original clip (resolution = 1920x1080, CRF = 22). The second was encoded either at one of the two standard variant settings appropriate to this session (2/3rds of trials) or was a repeat of the identical minimally-impaired clip (1/3rd of trials). These repeated trials served as "hidden references" and were subsequently used to calculate a differential DS-MOS score ([2] and see below). This led to a total of 90 trials, which were presented in a random order to each rater. On each trial, the rater was asked "Compared to the first video, the difference in quality of the second video was:" and provided with a 5-point response scale labelled "Very annoying," "Annoying," "Slightly annoying," "Visible but not annoying," and "Not visible." Twenty-two raters took part in the first session, and 24 in the second session (22 raters took part in both sessions). Responses from each rater were screened to check for random or non-discriminating responses, and data from three raters were excluded from the analysis due to non-discriminating responses (over 95% of all responses were of a single rating, across all variant types). For analysis, each rater's rating of each standard variant video clip was normalised relative to their rating for that same clip when presented with minimal impairment as a hidden reference. This yielded a Differential DS-MOS score, in which a score of "5" indicates that that variant was judged to be of as high a quality as its minimally impaired version.

*Adaptive Paired Comparisons method:* On each trial, a pair of ten-second clips was presented. One of the videos was a "standard" video encoded at one of the two standard variant settings appropriate to this session. The other was a "reference" version of that same video clip, encoded at one of the 50 quality levels of the reference scale. The level of the reference presented on each trial was determined by an active learning procedure described above, based on the rater's responses on previous trials involving that standard variant. The order of the standard and reference was random on each trial, and the rater reported whether the first or the second video had been of higher visual quality. Adaptive procedures terminated after 30 trials (one trial per video clip), and video clips were randomly assigned to the 60 trials within each session. For analysis, each rater's response data for each standard variant were expressed as the proportion of times the rater reported that the reference video looked better, as a function of the reference quality level. A logistic function with variable midpoint, slope, and upper and lower asymptotes was fitted to these data using non-linear least squares, and the midpoint of the fitted function was taken as the estimate of subjective quality for this variant for this rater. Twenty-two raters took part in the first session, and 24 in the second session. For each variant, some raters' data were too poorly discriminative to be able to fit; this led to the exclusion of 3, 4, 0 and 3 raters' data from the 180p, 360p (first session), 360p (second session) and 720p variant conditions, respectively.

V. RESULTS

Figure 3 shows subjective quality of the same 30 video clips estimated via the three methods: APC, MOS and DS-MOS using the test procedure described in IV.B. Since each method delivers estimates on different scales, quality estimates cannot be directly compared. Instead, we can compare the goodness of these methods by looking at the effect sizes of key comparisons between video variants. The effect size is a measure of the size of the difference of interest relative to the variability within conditions [8]. A good measurement method is precise (i.e. small variance in estimates across raters for the same variant, leading to larger effect sizes between different variants, and therefore greater statistical sensitivity to small differences in perceived quality), and reliable (i.e. repeated measurements of the same physical stimuli deliver highly similar quality estimates). Effect sizes for key comparisons are shown in Table 1. Key findings are:

1. The APC method was superior to the DS-MOS method in that it yielded larger effect sizes for both of the true-difference comparisons (180p vs 360p and 360p vs 720p), and a smaller effect size for the no-physical-difference comparison (360p across the two sessions).
2. Both APC and DS-MOS were superior to MOS in that they yielded larger effect sizes for the true-difference comparisons, and smaller effect sizes for the no-physical-difference comparison.
3. MOS alone returned a false-positive, concluding that there is a statistically significant difference in quality between the physically identical 360p videos measured on different days and in different contexts (i.e. MOS has poor test-retest reliability under these conditions).

Additional reasons to prefer APC are that (a) finer gradations of quality judgement are possible within the upper and lower range of the scales than with methods involving verbal description - e.g., by providing exceptionally high quality videos at the high end of the reference scale, raters can finely titrate the subjective quality of videos that could all be broadly described as "excellent", and (b) accurate and reliable results can be obtained with fewer raters, and with non-expert raters.

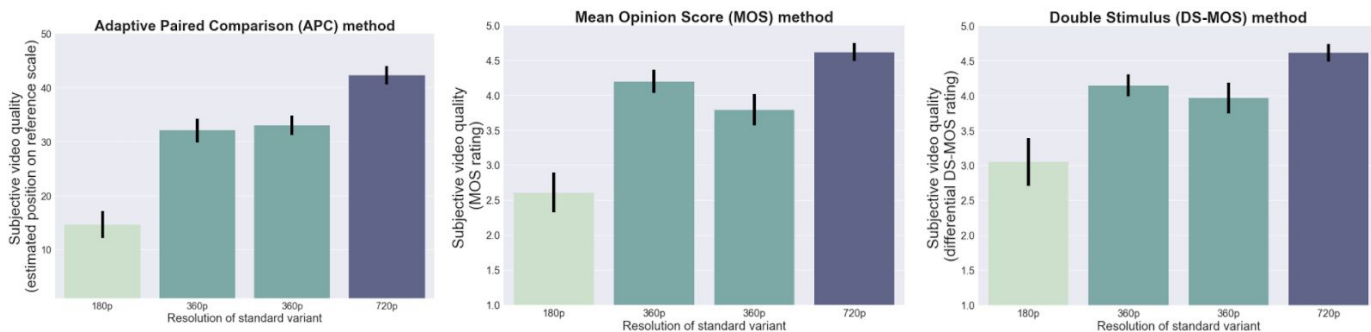

Figure 3. Mean subjective quality for three variant resolutions, measured by (left) our Adaptive Paired Comparison method, (middle) Mean Opinion Score, and (right) Double-Stimulus MOS. Error bars indicate 95% confidence intervals, calculated across raters.

|  | Effect size under each measurement method | | |
|---|---|---|---|
| Variants compared | MOS | DS-MOS | APC |
| 180p vs 360p | 2.80* | 1.28* | 4.50* |
| 360p vs 720p | 1.92* | 1.31* | 2.70* |
| 360p vs 360p (test-retest reliability) | 0.89* | 0.38 | 0.06 |

TABLE I. EFFECT SIZE FOR EACH KEY COMPARISON BETWEEN METHODS. Repeated measures $d$ as defined in [8]. Asterisks indicate that the subjective quality of variants was significantly different according to this method ($p < 0.05$, Bonferroni corrected for multiple comparisons).

## VI. DISCUSSION

Adaptive Paired Comparisons testing is superior to opinion-score-based methods for measuring subjective quality under weakly-controlled conditions, such as non-experts watching content on mobile devices under naturalistic viewing conditions. Here, the lack of tight control within and between participants, and lack of regularly reinforced anchor-points for the five quality evaluation levels means that MOS data tend to be noisy, and are biased by the context of other videos presented. Paired-comparison tasks compensate for some of these weaknesses by containing an objective reference stimulus on every trial, requiring of the observer only a *relative* quality judgement. Adaptive stimulus sampling reduces the number of trials required to a manageable number.

### A. Future Directions

The APC method can easily be extended to estimating properties of the psychometric function other than the midpoint, for example if the experimenter is interested in the sensitivity of observers to gradations of quality, they may directly estimate the slope as well as the midpoint. Additionally, individual participants could be treated as samples from an underlying population psychometric function, so that responses not only from earlier within the session, but from previous sessions with different individuals, are used to inform the selection of the next stimulus.

The method also lends itself to learning automated perceptual metrics [e.g. 9] more reliably by providing more precise and unbiased human quality judgements as training data. In conclusion, APC is the preferred method for measuring subjective video or image quality assessments, especially under weakly controlled viewing conditions.

ACKNOWLEDGEMENTS

The authors would like to thank the Magic Pony and HCOMP teams at Twitter for their effort and contributions.

REFERENCES

[1] Zhou Wang, A. C. Bovik, H. R. Sheikh and E. P. Simoncelli, "Image quality assessment: from error visibility to structural similarity," in *IEEE Transactions on Image Processing*, vol. 13, no. 4, pp. 600-612, 2004.
[2] Aaron, A., Li, Z., Manohara, M., De Cock, J. and Ronca, D, "Per-Title Encode Optimization," in online blogpost https://medium.com/netflix-techblog/per-title-encode-optimization-7e99442b62a2, Dec. 14, 2015.
[3] Methodology for the subjective assessment of the quality of television pictures, Rec. ITU-R BT. 500 -13, Jan. 2012.
[4] Subjective assessment of video quality using Expert Viewing Protocol, ITU-R BT.2095, June 2017
[5] Kingdom, F.A.A., and Prins, N., *Psychophysics: A Practical Introduction*, Elsevier Academic Press, London, 2010.
[6] Leek, M.R. "Adaptive procedures in psychophysical research," *Perception & Psychophysics*, vol. 63, no. 8, pp. 1279-1292, 2001.
[7] Huszár, F., and Houlsby, N.M.T, "Adaptive Bayesian quantum tomography," *Physical Review A*, vol. 85, no. 5:052120, pp. 1-5, 2012.
[8] Cohen, J. *Statistical Power Analysis for the Behavioral Sciences*, Routledge, 1988.
[9] Talebi, H., and Milanfar, P., "Learned perceptual image enhancement," *arXiv*, https://arxiv.org/abs/1712.02864, Dec. 7, 2017.